\documentclass{iopart}

\usepackage{graphics}
\usepackage{epsf}
\usepackage{epsfig}

\begin{document} 

\flushbottom

\title[Mean-field scaling function of absorbing phase transitions
with a conserved field]{Mean-field scaling function of the
universality class of absorbing phase transitions 
with a conserved field}


\author{S. L\"ubeck$^\dagger$ and A. Hucht$^\ddagger$}

\address{
$^\dagger$Weizmann Institute, Department of Physics of
Complex Systems,\\ 76100 Rehovot, Israel}
\address{
$^\ddagger$Institut f\"ur Physik, 
Gerhard-Mercator-Universit\"at,\\
47048 Duisburg, Germany\\[2mm] 
sven.lubeck@weizmann.ac.il, fred@thp.uni-duisburg.de\\[2mm]
Received 6 March 2002
}


{\vspace{-7.0cm}
\noindent
{accepted for publication in \it J.~Phys.~A (2002)}
\vspace{6.5cm}}





\begin{abstract}
We consider two mean-field like models which belong to the
universality class of absorbing phase transitions
with a conserved field.
In both cases we derive analytically the
order parameter as function of the control parameter
and of an external field conjugated to the
order parameter.
This allows us to calculate the universal
scaling function of the mean-field behavior.
The obtained universal function is in 
perfect agreement with recently obtained
numerical data of the corresponding
five and six dimensional models,
showing that four is the upper critical dimension
of this particular universality class.
\end{abstract}




\section{Introduction}

The scaling behavior of directed percolation is recognized 
as the paradigmatic example of the critical behavior
of several non-equilibrium systems which exhibits a 
continuous phase transition from an active state
to an absorbing non-active state 
(see for instance~\cite{GRINSTEIN_2,HINRICHSEN_1}).
The widespread occurrence of such systems in
physics, biology, as well as catalytic chemical 
reactions is reflected by the well known universality
hypothesis of Janssen and Grassberger that models which 
exhibit a continuous phase transition to a single
absorbing state generally belong to the universality class
of directed percolation~\cite{JANSSEN_1,GRASSBERGER_2}.
Introducing an additional symmetry the critical
behavior differs from directed percolation.
In particular particle conservation leads to
a new universality class of absorbing phase transitions
with a conserved field as pointed out in~\cite{ROSSI_1}.
In this work the authors introduced two models,
the conserved lattice gas (CLG) as well as a 
conserved threshold transfer process (CTTP).
The latter one is a conserved modification
of the threshold transfer process introduced
in~\cite{MENDES_1}.
Both models display a continuous phase transition
from an active to an inactive phase.
The density of active sites $\rho_{\rm a}$ is the 
order parameter of the phase transition controlled
by the total density of particles $\rho$, i.e., 
$\rho_{\rm a}>0$ if the density exceeds the
critical value $\rho_{\rm c}$ and zero otherwise. 
As usual in second order phase transitions the
order parameter vanishes algebraically at the
transition point.
The corresponding order parameter exponent as well as the 
exponent of the order parameter fluctuations of the
CLG are determined in~\cite{LUEB_19} for various
dimensions.

The scaling behavior of the CLG model in 
an external field conjugated to the order parameter
was considered recently~\cite{LUEB_22}.
The external field is realized by movements
of inactive particles which may be activated in this way.
Thus the field creates active particles without
violating the particle conservation.
Taking into account this additional scaling field 
the order parameter obeys the scaling ansatz
\begin{equation}
\rho_{\rm{a}}(\delta\rho, h) \; \sim \; 
\lambda\, \, {\tilde r}
(\delta \rho \; \lambda^{-1/\beta}, h \; \lambda^{-\sigma/\beta})
\label{eq:scal_ansatz}
\end{equation}
with the critical exponents $\beta$ and $\sigma$,
the scaling function~${\tilde r}$, the reduced control
parameter $\delta\rho=\rho/\rho_{\rm c}-1$,
and the external field~$h$.
Choosing $\delta\rho \, \lambda^{-1/\beta}=1$ one gets
for zero fields $\rho_{\rm a} \sim {\tilde r}(1,0) \,\delta\rho^\beta$,
whereas $h \, \lambda^{-\sigma / \beta}=1$ leads 
at the critical density to 
$\rho_{\rm a} \sim {\tilde r}(0,1) h^{\beta/\sigma}$.
Except of the critical point $(\delta\rho=0, h=0)$ the 
scaling function ${\tilde r}(x,y)$ is smooth and analytic
but it is not universal since it may depend, like the 
value of $\rho_{\rm c}$, on the details of the considered 
systems (here e.g.~the lattice structure, the update scheme, etc.).

A universal scaling function ${\tilde R}$ can be introduced
if one allows non-universal metric factors $c_i$ 
for the scaling arguments $\delta\rho$ and $h$
(see for instance~\cite{PRIVMAN_1}), i.e.,
\begin{equation}
\rho_{\rm{a}}(\delta\rho, h) \; \sim \; 
\lambda\, \, {\tilde R}
(c_1 \, \delta \rho \; \lambda^{-1/\beta}, c_2\, h \; \lambda^{-\sigma/\beta}),
\label{eq:uni_scal_ansatz}
\end{equation} 
and the scaling function is normed by
the conditions ${\tilde R}(1,0)={\tilde R}(0,1)=1$.
Then the function ${\tilde R}(x,y) $ is universal, i.e.,
similar to the critical exponents  
${\tilde R}(x,y)$ is identical for all models 
which belong to the same universality class.
But the non-universal metric factors differ 
again between the models and may depend on the lattice 
structure, the used update scheme etc. 

The non-universal metric factors can be easily
determined by the scaling behavior of the order
parameter at zero field and at the critical density,
respectively.
Choosing $c_1 \, \delta\rho \, \lambda^{-1/\beta}=1$ one gets
for zero fields ($h=0$)
\begin{equation}
\rho_{\rm a}(\delta\rho,0) \; \sim \; (c_1 \, \delta\rho )^\beta
\label{eq:uni_scal_ord_01}
\end{equation}
whereas $c_2 \, h \, \lambda^{-\sigma / \beta}=1$ leads 
at the critical density ($\delta\rho=0$) to 
\begin{equation}
\rho_{\rm a}(0,h) \; \sim \; (c_2 \, h )^{\beta/\sigma}.
\label{eq:uni_scal_ord_02}
\end{equation}

In this work we derive the universal scaling 
function~${\tilde R}$ of the mean-field solution
of the universality class of absorbing phase 
transitions with a conserved field.
In particular we consider analytically the CLG and the CTTP
with particle hopping to randomly chosen sites
on the whole lattice. 
This unrestricted particle hopping breaks long
range correlations and the scaling behavior is 
characterized by the mean-field exponents 
(see~\cite{LUEB_20}).
Neglecting correlations it is possible to derive
analytically the order parameter as a function of 
the control parameter and of the external field.
The obtained universal function is in perfect
agreement with recently obtained numerical data
of the five and six dimensional CLG and CTTP 
in an external field.

\section{The conserved lattice gas}

We consider the CLG model on a chain with $L$ sites and 
periodic boundary conditions.
At the beginning one distributes randomly $N=\rho L$ 
particles on the system where $\rho$ denotes the
particle density.
A particle is called active if at least one of its two
neighboring sites is occupied.
In the original CLG model active particles jump in
the next update step to one of their empty nearest 
neighbor site, selected at random~\cite{ROSSI_1}.
In the steady state the system is characterized by the
density of active sites $\rho_{\rm a}$ which
depends on $\rho$.
The density of inactive sites is given by 
$\rho_{\rm i}=\rho-\rho_{\rm a}$ 
and $1-\rho$ is the density of empty sites.

\begin{table}[t]
\caption{The configuration of a CLG lattice before ($\cal{C}$) and
after ($\cal{C'}$) a particle hopping.
Only the target lattice site where a particle hops onto and its
left and right neighboring sites are shown.
Empty sites are marked by $\circ$, inactive sites are
marked by  $\ast$, and active sites by $\bullet$.
$\Delta n$ denotes the change of the number of active
sites due to the particle hopping and $p$ is the
corresponding probability of the configuration $\cal{C}$
if one neglects spatial correlations.}
\label{table:clg_conf}
\begin{indented}
\item[]
\begin{tabular}{ccrl}
\br
$\cal{C}$ & $\cal{C'}$       & $\Delta n$       & $p(\cal{C} \to \cal{C}^\prime)$ \\  
\mr
$\circ \circ \circ $     &  $\circ \ast \circ $         & $-1$ & $\rho_{\rm a} \; (1-\rho) \; (1-\rho)^2$   \\ 
$\ast  \circ \circ$      &  $\bullet \bullet \circ$     & $+1$ & $\rho_{\rm a} \; (1-\rho) \; 2 \rho_{\rm i}  (1-\rho)$   \\ 
$\ast  \circ \ast $      &  $\bullet \bullet \bullet$   & $+2$ & $\rho_{\rm a} \; (1-\rho) \; \rho_{\rm i}^2$   \\ 
$\bullet \circ \circ$    &  $\bullet \bullet \circ$     & $0$  & $\rho_{\rm a} \; (1-\rho) \; 2 \rho_{\rm a} (1-\rho)$   \\ 
$\bullet \circ \bullet$  &  $\bullet \bullet \bullet$   & $0$  & $\rho_{\rm a} \; (1-\rho) \; \rho_{\rm a}^2$   \\ 
$\bullet \circ \ast $    &  $\bullet \bullet \bullet$   & $+1$ & $\rho_{\rm a} \; (1-\rho) \; 2 \rho_{\rm a} \rho_{\rm i}$   \\ 
\br
\end{tabular}
\end{indented}
\end{table}

We introduced in~\cite{LUEB_20} a modification of the 
CLG model where active particles 
are moved to a randomly chosen empty lattice site
which suppresses long range correlations.
A given lattice site is active with a probability
$\rho_{\rm a}$ and with the probability $1-\rho$ 
it may be moved to an empty lattice site.
Depending on the neighborhood of this new lattice
site the number of active sites may change. 
For instance if both new neighbors of the moved
particle are empty the number of active particles
is reduced by one, $\Delta n =-1$.
Without correlations the corresponding probability
for this process is $\rho_{\rm a}\, (1-\rho)^3$.
In the case that one of the new neighbors of the 
moved particle is occupied by an inactive particle 
($\rho_{\rm i}$)
and the second neighbor is empty ($1-\rho$), 
the number of active sites is increased by one ($\Delta n =1$). 
The corresponding probability is given by 
$p=2 \rho_{\rm a} \, \rho_{\rm i} \, (1-\rho)^2$.
All other possible configurations and the corresponding
probabilities are listed in table\,\ref{table:clg_conf}.

The probabilities that the number of active particles 
are changed by $\Delta n$ are given by 
\begin{equation}
\begin{array}{lcl}
\label{eq:clg_prob_non_field}
p_{\scriptscriptstyle \Delta n = -1} & =  & (1-\rho)\; \rho_{\rm a} \; (1-\rho)^2 ,\\ \\
p_{\scriptscriptstyle \Delta n = 0}  & =  & (1-\rho)\; \rho_{\rm a} \; [2 \rho_{\rm a}(1-\rho)+\rho_{\rm a}^2] , \\ \\
p_{\scriptscriptstyle \Delta n = 1 } & =  & (1-\rho)\; \rho_{\rm a} \; [2 \rho_{\rm i} (1-\rho) + 2 \rho_{\rm a} \rho_{\rm i}] , \\ \\
p_{\scriptscriptstyle \Delta n = 2 } & =  & (1-\rho)\; \rho_{\rm a} \; \rho_{\rm i}^2 .\\
\end{array}
\end{equation}
The expectation value of $\Delta n$ is 
\begin{equation}
E[\Delta n] \; = \; \sum_{\Delta n=-1}^{2} \, \Delta n \, \,
p_{\scriptscriptstyle \Delta n} 
\; = \; (1-\rho) \, \rho_{\rm a} \, [-1 -2\rho_{\rm a} + 4 \rho -\rho^2].
\label{eq:clg_expect_value}
\end{equation}
As pointed out in~\cite{LUEB_20}
the average number of active sites is constant in the
stationary state, i.e., the expectation value
of $\Delta n$ should be zero in the steady state.
Using the constraint $E[\Delta n]=0$ one gets
\begin{equation}
\rho=1 \quad \vee \quad \rho_{\rm a}=0 \quad \vee \quad 
-1 -2\rho_{\rm a} + 4 \rho -\rho^2=0.
\end{equation}
The first equation corresponds to a system were all sites
are occupied ($\rho_{\rm a}=1$) and no dynamics can take 
place whereas the
absorbing state is reflected by the second equation.
The non-trivial third equation corresponds
for $\rho_{\rm a}>0$ to the active phase and one
gets for the order parameter in leading order
\begin{equation}
\rho_{\rm a} \;  = \; \frac{\,4 \rho- \rho^2 -1\,}{2}
\; = \; (2\sqrt{3}-3) \, \delta\rho \; + \; \Or(\delta\rho^2)
\label{eq:clg_rho_a}
\end{equation}
with the critical density $\rho_{\rm c}=2-\sqrt{3}$~\cite{LUEB_20}.
Thus we have obtained the critical exponent $\beta=1$
as well as the non-universal metric factor $c_1=2 \sqrt{3} -3$.

\begin{figure}[t]
  \includegraphics[width=8.6cm,angle=0]{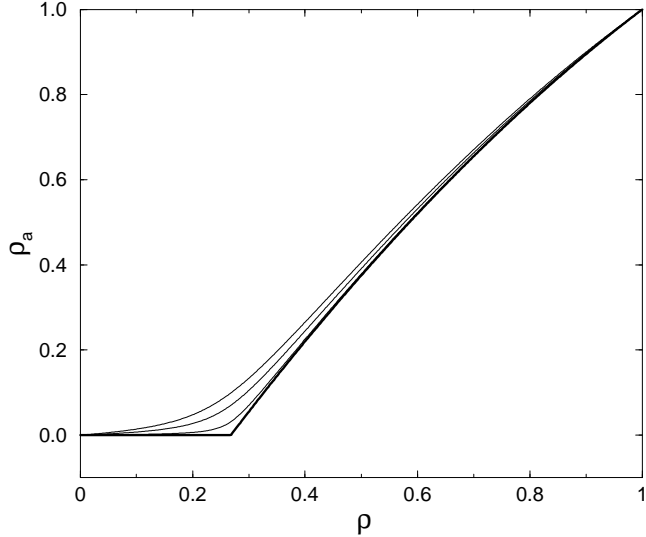}
  \caption{
    The order parameter of the CLG model as a function of
    the particle density~$\rho$ and the applied external field~$h$
    [see equation\,(\protect\ref{eq:clg_ord_field})].
    The thick line corresponds to $h=0$ whereas the thin lines
    correspond to $h=0.1, 0.05, 0.01$ (from top to bottom).
   }
  \label{fig:rho_a_clg_sketch} 
\end{figure}

In the case that an external field is applied
non-active sites may be activated (see~\cite{LUEB_22}).
The probability that a site is occupied and has two
empty neighbors is $\rho (1-\rho)^2$.
These particles are activated with probability~$h$,
where $h$ denotes
the strength of the applied field.
In this process the number of active sites is increased
($\Delta n=1$) and the probability $p_{\Delta n=1}$ 
is modified to be
\begin{equation}
p_{\scriptscriptstyle \Delta n = 1 } 
\; = \;   (1-\rho)\; \rho_{\rm a} \; 
[2 (1-\rho) \rho_{\rm i}+2 \rho_{\rm i} \rho_{\rm a}]
\; + \; (1-\rho)^2 \, \rho \, h.
\label{eq:prob_field}
\end{equation}
Using again the steady state condition $E[\Delta n]=0$
one gets the equations
\begin{equation}
\rho=1 \quad \vee \quad 
\rho_{\rm a}\,[-1 -2\rho_{\rm a} + 4 \rho -\rho^2] \,+ \,
(1-\rho) \, \rho \, h \; = \; 0.
\end{equation}
The first equation corresponds again to the trivial case of
a totally occupied lattice whereas the second equation yields
the solutions
\begin{equation}
\rho_{\rm a} \; = \;
\frac{1}{4} \, \left (
-1 + 4 \rho  - \rho^2 \pm
\sqrt{8 h (1-\rho) \rho + (-1+ 4 \rho - \rho^2)^2}
\right ) .
\label{eq:clg_ord_field}
\end{equation}
The solution with the $+$~sign describes the order 
parameter $\rho_{\rm a}(\rho, h)$
as a function of the density and of the external field 
whereas the $-$~sign solution yields 
negative densities for the order parameter for all values 
of $\rho$ and $h$.
A sketch of the order parameter for various fields 
is presented in figure\,\ref{fig:rho_a_clg_sketch}.

At the critical density $\rho_{\rm c}=2-\sqrt{3}$ 
the order parameter is given by
\begin{equation}
\rho_{\rm a}(\rho_{\rm c},h) \; = \; 
\sqrt{ \, \frac{3 \sqrt{3}-5}{2} \,} \; \sqrt{h} 
\label{eq:clg_ord_scal_field}
\end{equation}
i.e., the field scaling exponent is $\sigma=2$
and the non-universal metric factor is
$c_2=(3\sqrt{3}-5)/2$.

In the following we derive the universal scaling
function $R(x,y)$ of the mean-field solution.
Therefore we write the order parameter [equation\,(\ref{eq:clg_ord_field})] 
as a function of the reduced control parameter~$\delta\rho$   
and consider the function
$\rho_{\rm a}(\delta \rho, h) / \sqrt{h}$.
Since we are interested in the scaling behavior in the
vicinity of the critical point we perform the 
limits $\rho_{\rm a}\to 0$, $\delta\rho\to 0$ and $h\to 0$
with the constraint that $\rho_{\rm a}/\sqrt{h}$ and 
$\delta\rho / \sqrt{h}$ are finite.
Thus all terms which scales as $\delta\rho^2/\sqrt{h}$ or 
$\delta\rho \, \sqrt{h}$ vanishes and in 
leading order and we get
\begin{equation}
\frac{\rho_{\rm a}(\delta \rho, h)}{\sqrt{h}} \; = \;
\frac{2\sqrt{3}-3}{2} \, \frac{\delta\rho}{\sqrt{h}} \; + \;
\sqrt{ \, \frac{3\sqrt{3}-5}{2} + 
\left ( \frac{2\sqrt{3}-3}{2} \, \frac{\delta\rho}{\sqrt{h}} 
\right )^2}.
\label{eq:clg_ord_par_uni_func}
\end{equation}
Introducing the non-universal metric factors $c_1=2\sqrt{3}-3$
and $c_2=(3 \sqrt{3}-5)/2$ one gets the universal function
\begin{equation}
{\tilde R}(c_1 \delta\rho,c_2 h)
\;=\;
\frac{c_1\delta\rho}{2} \; + \;
\sqrt{\, c_2 h + \left (\frac{c_1 \delta\rho}{2}\right )^2\;}.
\label{eq:ord_par_uni_func}
\end{equation}
Equations\,(\ref{eq:clg_rho_a},\ref{eq:clg_ord_scal_field}) are
recovered from this result by 
setting $h=0$ and $\delta\rho = 0$, respectively.
Furthermore we get 
${\tilde R}(1,0)={\tilde R}(0,1)=1$
as required above.

As usual in scaling analysis 
(see for instance~\cite{LUEB_22}) the order parameter
as well as the control parameter are rescaled 
by the field in order to obtain a data collapse
[setting $c_2\,h\,\lambda^{-\sigma/\beta}=1$ in 
equation\,(\ref{eq:uni_scal_ansatz})].
In this case one gets the universal function
\begin{equation}
\frac{\rho_{\rm a}(\delta\rho, h)}{\sqrt{c_2 h}} \; \sim \; 
{\tilde R}(x,1) \; = \; 
\frac{x}{2}  \, +\, \sqrt{\,1 \, + \, \left ( \frac{x}{2} \right )^2\,}
\label{eq:uni_func_scal_plot}
\end{equation}
where the scaling argument is given by 
$x=c_1 \delta\rho / \sqrt{c_2 h}$.

For the sake of simplicity we derived the scaling function
of the one-dimensional CLG model only.
A straight forward extension to higher dimensional systems for $h=0$
was already presented in~\cite{LUEB_20}.
The increased number of nearest neighbours in higher 
dimensions affects the non-universal quantities
$\rho_{\rm c}$, $c_1$, and $c_2$ only, but not the
critical exponents and the universal scaling function.

\section{The conserved threshold transfer process}

A similar analysis can be performed for the
CTTP with random neighbor hopping.
In the CTTP lattice sites may be empty, occupied or
double occupied.
Double occupied lattice sites are considered as
active and one tries to transfer both particles of 
each active site to randomly chosen lattice sites.
Recently performed numerical investigations in 
dimensions $d=2,3,4,5,6$ confirm the conjecture 
of~\cite{ROSSI_1} that the CLG and the CTTP belong
to the same universality class~\cite{LUEB_24}.
Analogous to the above presented analysis we derive
the mean-field critical behavior of the order parameter of the
CTTP with random neighbor hopping.

In the following we denote the densities of sites with
$\rho_{\rm a}$ (double occupied and active),
$\rho_{\rm i}$ (single occupied and inactive),
and 
$\rho_{\rm e}$ (empty). 
Normalization requires $\rho_{\rm e} + \rho_{\rm i} + \rho_{\rm a}=1$ and
the particle conservation is reflected by the
equation $\rho_{\rm i} + 2 \rho_{\rm a}=\rho$ where the control 
parameter $\rho$ denotes again the density of particles on 
a $D$-dimensional lattice, i.e., $\rho=N/L^D$.
The probability that a given lattice site $s$ is active is
therefore $\rho_{\rm a}$.
In this case the two active particles are tried to transfer 
to two randomly chosen lattice sites $t_1$ and $t_2$.
In the case that both sites are empty the two particles
are moved to the empty sites and the number of
active sites is decreased by one ($\Delta n=-1$).
The probability for this process is $\rho_{\rm a} \, \rho_{\rm e}^2$.
All other possible configurations and the corresponding
probabilities are listed in table\,\ref{table:cttp_conf}.
The probabilities that the number of active particles 
are changed by $\Delta n$ are thus given by 
\begin{equation}
\begin{array}{lcl}
\label{eq:cttp_prob_non_field}
p_{\scriptscriptstyle \Delta n = -1} & =  & \rho_{\rm a} \; [\rho_{\rm e}^2 \, + \, 2 \,\rho_{\rm e} \, \rho_{\rm a}] \\ \\
p_{\scriptscriptstyle \Delta n = 0}  & =  & \rho_{\rm a} \; [2\, \rho_{\rm e} \rho_{\rm i}+ 2 \rho_{\rm i} \rho_{\rm a} + \rho_{\rm a}^2] \\ \\
p_{\scriptscriptstyle \Delta n = 1 } & =  & \rho_{\rm a} \; \rho_{\rm i}^2.\\ \\
\end{array}
\end{equation}
The steady state condition $E[\Delta n]=0$ leads to the
equations 
\begin{equation}
\rho_{\rm a}=0 \quad \vee \quad 
-1 + 2 \rho  -4 \rho_{\rm a} +  \rho_{\rm a}^2 =0.
\end{equation}   
Again the first equation corresponds to the absorbing
state and the second equation yields the order parameter
as a function of the particle density 
\begin{equation}
\rho_{\rm a} \; = \; 2 \pm \sqrt{5-2\rho\,}.
\label{eq:cttp_ord_par_01}
\end{equation}
Here, the $+$ solution can be neglected ($\rho_{\rm a}>1$) and the
$-$ solution describes the order parameter behavior
above the critical density $\rho_{\rm c} =1/2$.
Close to this critical point the order parameter
scales in leading order as
\begin{equation}
\rho_{\rm a} \; = \; \frac{1}{4} \,\delta\rho \, + \, \Or (\delta\rho^2),
\label{eq:cttp_ord_par_02}
\end{equation}
i.e., the non-universal metric factor of the CTTP is 
$c_1=1/4$ and the critical exponent is in agreement with
the CLG model $\beta=1$.

\begin{table}[t]
\caption{The configuration of a CTTP lattice before ($s,t_1,t_2$) and
after ($s^{\prime},t_1^{\prime},t_2^{\prime}$) a particle hopping.
Only the source lattice site ($s$) and its
two targets sites ($t_1$ and $t_2$) where the two particles
may be moved are shown.
$\Delta n$ denotes the change of the number of active
sites due to the particle hopping and $p$ is the
corresponding probability of the configuration ($s,t_1,t_2$)
if one neglects spatial correlations.}
\label{table:cttp_conf}
\begin{indented}
\item[]
\begin{tabular}{ccccccll}
\br
$\;s\;$ &  $\;t_1\;$ & $\;t_2\;$ & $\;s^{\prime}\;$ & $\;t_1^{\prime}\;$ & $\;t_2^{\prime}\;$  & $\Delta n\;\;\;$       & $\;p(s,t_1,t_2)\;$ \\  
\mr
$2$ &  $0$   & $0$   & $0$          & $1$            & $1$             & $-1$             & $\rho_{\rm a} \, \rho_{\rm e}^2$    \\ 
$2$ &  $0$   & $1$   & $0$          & $1$            & $2$             & $0$              & $\rho_{\rm a} \, 2\, \rho_{\rm e} \, \rho_{\rm i}$   \\ 
$2$ &  $0$   & $2$   & $1$          & $1$            & $2$             & $-1$             & $\rho_{\rm a} \, 2\, \rho_{\rm e} \, \rho_{\rm a}$   \\ 
$2$ &  $1$   & $1$   & $0$          & $2$            & $2$             & $+1$             & $\rho_{\rm a} \, \rho_{\rm i}^2$    \\ 
$2$ &  $1$   & $2$   & $1$          & $2$            & $2$             & $0$              & $\rho_{\rm a} \, 2\, \rho_{\rm i} \, \rho_{\rm a}$   \\ 
$2$ &  $2$   & $2$   & $2$          & $2$            & $2$             & $0$              & $\rho_{\rm a} \, \rho_{\rm a}^2$   \\ 
\br
\end{tabular}
\end{indented}
\end{table}

Similar to the CLG model we now apply an external
field which activates single occupied sites.
The probability that the external field~$h$ acts to
a given site is $\rho_{\rm i} \, h$
and one tries to transfer this particle to a 
randomly chosen lattice site.
In the case that the activated particle is moved
to an empty lattice site the number of active 
site is unchanged by this field induced process ($\Delta n=0$).
The number of active sites is increased only if the
particle is moved to a single occupied 
lattice site ($\Delta n=+1$). 
The probability for this process is $\rho_{\rm i}^2 \, h$.
In order to incorporate the external field into the 
dynamics one has to modify $p_{\scriptscriptstyle \Delta n = 1 }$
accordingly and the steady state condition yields 
\begin{equation}
\rho_{\rm a} \, (-1 + 2 \rho - 4 \rho_{\rm a} + \rho_{\rm a}^2) \,+ \, h\,(\rho-2 \rho_{\rm a})^2
\; = \; 0.
\label{eq:cttp_ord_par_field_01}
\end{equation}
At the critical density $\rho_{\rm c}=1/2$ the
order parameter scales with the external field
according to 
\begin{equation}
\rho_{\rm a} \; = \; \frac{1}{4} \, h^{1/2} + \Or (h),
\label{eq:cttp_ord_par_field_02}
\end{equation} 
i.e., the critical exponent is again $\sigma=2$ and
the non-universal metric factor of the CTTP is
given by $c_2=1/16$.

In order to obtain the universal scaling 
function~${\tilde R}$ we set $\rho = \rho_c + \rho_c \, \delta\rho$
and transform equation\,(\ref{eq:cttp_ord_par_field_01})
to
\begin{equation}
\frac{\rho_{\rm a}}{\sqrt{h}} \, \left ( \frac{\delta\rho}{\sqrt{h}} \, - \,
4 \, \frac{\rho_{\rm a}}{\sqrt{h}} \, + \, \frac{\rho_{\rm a}^2}{\sqrt{h}}
\right )
\; + \; \left ( \frac{1}{2} + \frac{1}{2} \, \delta\rho - 2 \,\rho_{\rm a}^2
\right )^2
\; = \; 0 .
\label{eq:cttp_ord_par_field_03}
\end{equation}
Focusing to the critical scaling behavior ($h\to 0$, $\delta\rho\to 0$, 
$\rho_{\rm a}\to 0$ where again $\rho_{\rm a}/\sqrt{h}$ as well as $\delta\rho/\sqrt{h}$ 
is kept constant) we can neglect all 
irrelevant terms and get in leading order
\begin{equation}
\frac{\rho_{\rm a}}{\sqrt{h}} \, \left ( \frac{\delta\rho}{\sqrt{h}} \, - \,
4 \, \frac{\rho_{\rm a}}{\sqrt{h}} 
\right )
\; + \; \frac{1}{4}
\; = \; 0 .
\label{eq:cttp_ord_par_field_04}
\end{equation}
This equation can be easily solved and one gets 
\begin{equation}
\rho_{\rm a}(\delta\rho,h) \; = \; \frac{\delta\rho}{8}
\, \pm \, \sqrt{\frac{h}{16}\, + \, \left ( \frac{\delta\rho}{8} \right )^2}
\label{eq:cttp_ord_par_field_05}
\end{equation}
where the $-$~sign can be neglected since it yields negative
values of the order parameter.
Using the non-universal metric factor $c_1=1/4$ and $c_2=1/16$
we get eventually again equation\,(\ref{eq:ord_par_uni_func}),
i.e. both models the CLG as well as the CTTP are 
characterized by the same universal function ${\tilde R}(x,y)$
in the mean-field solution.
Furthermore, the obtained universal function ${\tilde R}$ 
agrees with that of the mean-field solution of directed
percolation (see for instance~\cite{JANSSEN_2}), i.e., 
although the CLG and CTTP differ from the directed percolation
scaling behavior in low dimensions they coincide on the 
mean-field level.

\section{Numerical simulations}

\begin{figure}[t]
  \includegraphics[width=8.6cm,angle=0]{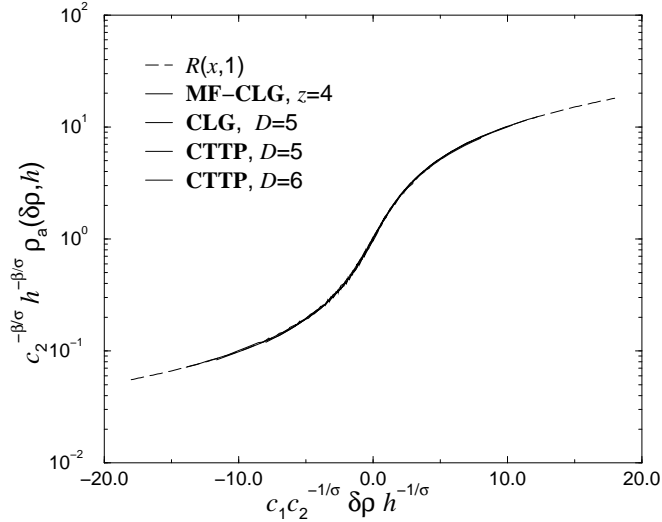}
  \caption{
    The mean-field universal function ${\tilde R}(x,1)$ 
    [see equation\,(\protect\ref{eq:uni_func_scal_plot})]
    of the universality of absorbing phase transitions
    with a conserved field.
    The numerical data of the five and six dimensional
    models are obtained from~\protect\cite{LUEB_22,LUEB_24}.
    Additionally we plot the data of a (mean-field like) CLG
    model with random neighbor hopping on a square lattice 
    ($z=4$ next neighbors) which was introduced in~\cite{LUEB_19}. 
    At least four different field values are plotted for each
    model.
   }
  \label{fig:uni_func_scal_plot} 
\end{figure}

In the following we compare our results
with those obtained from numerical simulations.
The upper critical dimension of the universality
class of absorbing phase transitions with a
conserved field is $D_{\rm c}=4$~\cite{LUEB_19}.
Thus we compare our results with the scaling
behavior of the CLG in $D=5$~\cite{LUEB_22}, the
CTTP in $D=5$ and $D=6$~\cite{LUEB_24}, as
well as with the scaling behavior of a two-dimensional
CLG on a square lattice where active particles are moved 
to randomly
chosen lattice sites~\cite{LUEB_20}.
In all models the order parameter is determined as a function
of the control parameter for various fields and the 
data are rescaled according to equation\,(\ref{eq:uni_func_scal_plot}).
Varying the non-universal metric factors we observe 
a data-collapse with the universal function ${\tilde R}(x,1)$.
The corresponding curves are presented in 
figure\,\ref{fig:uni_func_scal_plot}.
As one can see, all numerically obtained curves fits
well with the derived universal function.
Furthermore the perfect data collapse of the curves 
for different dimensions, as well as for a mean-field model
clearly confirms that four is the upper critical dimension.\\[5mm]

Notice that the mean-field behavior of the 
CTTP order parameter was recently considered 
in~\cite{DICKMAN_3}. Using a cluster approximation
method the authors obtained equation (\ref{eq:cttp_ord_par_01})
that describes the zero-field behavior of the
order parameter.

\ack{S.\,L\"ubeck wishes to thank the Minerva 
Foundation (Max Planck Gesellschaft) for financial support.}

\end{document}